# Speed-dependent ice bandings in freezing colloidal suspensions

Jiaxue You[1], Jincheng Wang[1], Lilin Wang[2], Zhijun Wang[1*], Junjie Li[1] and Xin Lin[1*]

1-State Key Laboratory of Solidification Processing, Northwestern Polytechnical University, Xi'an 710072, P. R. China

2-School of Materials Science and Engineering, Xi'an University of Technology, Xi'an 710048, P. R. China

**Abstract:** Formation mechanism of speed-dependent ice bandings in freezing colloidal suspensions, of significance in frost heaving and materials science, remains a mystery. With quantitative experiments, we propose a possible mechanism of speed-dependent ice bandings by focusing on the particle packing density and dynamic interface undercooling. The particle packing density ahead of the freezing interface decreases with increasing pulling speeds, attributed to the speed-dependent packing of particles. Through affecting the curvature undercooling of pore ices, the speed-dependent packing of particles can be used to explain speed-dependent thicknesses of ice bandings. The dynamic interface undercooling was obtained to explore curvature undercooling of pore ices and quantitative details of transient interface positions, speeds and undercooling were given. All the evidences imply that the speed-dependent particle packing and dynamic interface undercooling are responsible for speed-dependent ice bandings.

**Introduction**

Solidification of colloidal suspensions plays a central role in many fields, such as earth and planetary science, food engineering, materials science and microfluidics etc. [1-7]. A particularly important observation in this complex system is that colloidal suspensions do not freeze uniformly: the frozen phase (i.e. ice) becomes segregated, trapping bulk regions of colloid within [8, 9]. Ice banding is one of the typical morphological segregations, where segregated lenses of ice align perpendicular to the direction of the thermal gradient. These perpendicular segregated patterns are of great importance due to their effects on the dynamics of frost heaving [10-12] and mechanical properties of freeze casting porous materials [3, 13].

* Corresponding author. Tel.:86-29-88460650; fax: 86-29-88491484
 E-mail address: zhjwang@nwpu.edu.cn (Zhijun Wang), xlin@nwpu.edu.cn (Xin Lin)

Recently, Anderson et al. [9] discovered pulling-speed-dependent ice bandings by means of experiments with directional freezing colloidal dispersions. The pattern formation of ice bandings is largely different in varying pulling speeds. They found pore ices penetrated into close-packed particles and formed frozen fringe, coexistence of particles and ice in the interstice in a low pulling speed. In high pulling speed, the frozen fringe vanished. Meanwhile, the thicknesses of both ice bandings and particle bandings increase with the decrease of pulling speeds. These results broaden the evolution of patterns in freezing colloidal suspensions. Nevertheless, the potential reason of the speed-dependent pattern formation remains unknown. Therefore, it is still an unsolved and significant issue on the physical processes controlling the formation of speed-dependent ice banding.

In previous investigation, the packing density in front of the solid/liquid interface was assumed as 0.64 near the random close packing [14, 15]. However, it was found that the ice banding formation is greatly dependent on the pulling speed, but without clear physical scenario of the speed-dependent effect [9]. It can be supposed that the packing density of particles ahead of the freezing ice $\phi_p$ may vary with increasing pulling speeds [1]. Therefore, the packing density of particles should be one of the key factors in the speed-dependent ice bandings.

In this paper, we investigated the ice bandings using an experimental facility with high precision to in-situ reveal the dynamic interface evolution of ice banding. Two quantitative characteristics, particle packing density in front of the freezing interface and dynamic interface undercooling, are proposed to reveal the speed-dependent ice bandings.

**Experimental methods**

In our experiments, α−alumina powder (Wanjing New Material, Hangzhou, China, ⩾99.95% purity) with a documented average diameter of ∼80nm (with 90% particles ranging from 40−120 nm in size) and a density of 3.97 g cm$^{-3}$ was used to prepare the suspensions. The alumina nano-suspensions were prepared strictly by following Ref.[9], in order to suppress the solute effects and highlight the particle effects on the freezing behavior of water. The particles were charge stabilized by a solution of analytical-grade HCl. The final pH of suspensions we used was tested by pH paper and in the range of yellow color (pH 5~7). Thus the suspensions have a zeta potential $\zeta$ of around 50~70 mV according to Fig.1 of Ref.[9], which is in the region of stable dispersions (pH < 7.5, $\zeta$ >40 mV). The prepared suspensions were added into rectangular glass capillary sample cells (with a cross-section of 1 mm×0.05 mm) before freezing. The initial volume fraction of particles is $\phi_0$= 15.97% (wt%= 43). The Bridgman freezing setup and experimental procedure have been described in Ref.[16, 17]. In the setup, the thermal gradient is produced by two heating and cooling zones separated by a gap. Sample translation across the thermal gradient is provided by a servo-drives motor. Observation is achieved through an optical microscope stage with a charge-coupled device (CCD) camera. During directional freezing, the thermal gradient was measured as G=7.23K/cm. Images were recorded via a CCD camera with 2580×1944 sensitive elements on a time-lapse video recorder and further analyzed by the software image processing. The optical contrast between the areas of ice and the areas of particle suspension (frozen and unfrozen) was large. Therefore, a simple image analysis technique was employed, where a threshold brightness was set to distinguish ice and particle suspension. We used Image-Pro Plus 6.0 to get data from the experimental pictures. In all images presented, segregated ice is bright and the particle suspension is dark.

**Speed-dependent ice bandings**

In order to explore the relation between pulling speeds and packing density of particles in more details, different pulling speeds of V=0.80, 3.50, 8.22 and 16 μm/s, are applied. The corresponding ice-banding patterns of steady state are shown in Fig.1(a), (b), (c), and (d), respectively. The dynamic processes for the speed-dependent ice bandings are presented in Movies S1, S2, S3 and S4, respectively, (Supplementary Information). The visible ice (the white part in the picture) is the segregated ice (i.e., ice banding), while particle bandings (the black part in the picture)

**3 / 16**

also include some unfrozen water among the particles and this interstice water may develop into pore ice in a colder ambient according to the curvature undercooling of pore ice [10]. For the solidified region, it is obvious that the thicknesses of both particle bandings and ice bandings decrease with increasing pulling speeds, which has the same tendency of Anderson's experiments [9], as shown in Fig.1. In the following, we analyze the proportion of the segregated ice $\phi_i$ in Fig.1 to obtain the packing density of particles $\phi_p$ in the particle bandings. Using the speed-dependent particle packing density $\phi_p$ and growth of pore ice, we analyze the speed-dependent thicknesses of particle bandings and ice bandings.

In Fig.1, the proportion of the segregated ice $\phi_i$ has a distinct change with the varying pulling speeds. This implies that the amount of the interstice water $\phi_w$ in particle bandings changed with different pulling speeds according to

$$\phi_i + \phi_w = 1 - \phi_0. \tag{1}$$

where $\phi_0$ is initial volume fraction of particles. Before the final stage of solidification, the volume expansion of ice can be release into far field. Therefore, we temporarily ignore the density different of ice and water. The amount of the interstice water in particle bandings can be used to calculate the packing density of particles $\phi_p$ in the particle bandings according to

$$\phi_p = \frac{\phi_0}{\phi_0 + \phi_w}. \tag{2}$$

Figure 2 presents the variation of area fractions of segregated ice and particle bandings based on their corresponding areas in Fig.1. The area fraction can be used to represent the volume fraction due to the limited thickness ($l_t$=0.05 mm) of the sample cell and the excellent thermal diffusivity ($D_t$≈0.143·$10^{-6}$ $m^2$/s) along the direction of thickness [16]. The characteristic thermal diffusion speed along the direction of the thickness $V_t$ (=$D_t/l_t$ ≈2.86·$10^3$ μm/s) is around three orders of magnitude larger than the pulling speeds V. Therefore, in the direction of sample thickness, there is almost no segregation between ice and particles. In addition, we used other independent experiments of mono-layer particles in Fig. 3 of Ref.[1] to confirm the results of Fig.2.

In Fig.2, the black solid line shows the volume fraction of segregated ice $\phi_i$ dramatically decreases with increasing pulling speeds. This implies the amount of interstice water $\phi_w$ in the particle bandings increases according to Eq.(1). According to Eq.(2), increasing interstice water $\phi_w$ will lead to decreasing particle packing density $\phi_p$ in the particle bandings. Therefore, the particle packing density $\phi_p$ in the particle bandings decreases with increasing pulling speeds. On the other hand, the red dotted line shows the volume fraction of particle bandings ($\phi_0+\phi_w$) increases with increasing pulling speeds, which also implies the particle packing density $\phi_p$ in the particle bandings decreases with increasing pulling speeds according to Eq.(2).

This speed-dependent particle packing density can be explained by the speed-dependent entropic ordering of particles [1, 18-20] and also confirmed by other independent experiments of mono-layer particles in Fig. 3 of Ref.[1]. When pulling speed is small, particles have enough time to finish entropic ordering and become close-packed ahead of the freezing interface. When the pulling speed is large, particles have little time to finish entropic ordering and become loose-packed in front of the freezing interface. On the other hand, the freezing speed will probably change the particle packing density even if the amorphous glasses may form due to the particulate size polydispersity. In the experiments of metallic glasses (such as the alloy $Zr_{55}Cu_{30}Ni_5Al_{10}$ of atomic size polydispersity), high-frequency dynamic micropillar tests have been used to probe both atomic clusters and defects. It has been found that the atomic packing density will decrease as the freezing speed increases [21, 22]. And particles are usually seen as big atoms [18]. Therefore, packing density of particles $\phi_p$ in particle bandings probably decreases with increasing pulling speeds [1].

The speed-dependent packing density of particles combined with the growth of pore ice can reveal why the thicknesses of periodical ice bandings depend greatly on pulling speeds in a schematic Fig.3. The thickness of particle bandings $H_p$ can be calculated as the curvature undercooling of pore ice

$$H_p = \Delta T_R/G. \tag{3}$$

where $\Delta T_R = \frac{T_m \gamma}{\rho_i L_f} \times \frac{2}{R_p}$ is the curvature undercooling of pore ice [10]. $T_m$ is the melting point of ice, $\gamma$ is the ice/water interfacial energy. $\rho_i$ is the density of ice. $L_f$ is latent heat of fusion. $R_p = \frac{2(1-\phi_p)}{\rho_p A_p \phi_p}$ [23] is the effective pore radius determined by the interaction between the particles and the interface. $\rho_p$ and $A_p$ are the particle density and specific surface area, respectively. Therefore, the thickness of particle bandings $H_p$ has a relation with particle packing density $\phi_p$ as

$$H_p = \frac{T_m \gamma \rho_p A_p}{G \rho_i L_f} \times \frac{\phi_p}{1-\phi_p}. \tag{4}$$

When the pulling speed is small, the packing density of particles $\phi_p$ is large, leading to a large thickness of particle bandings $H_p$ according to Eq.(4). On the contrary, when the pulling speed is large, the packing density of particles $\phi_p$ is small, leading to a small thickness of particle bandings $H_p$. Therefore, the thickness of particle bandings decreases with increasing pulling speeds.

As to the speed-dependent thickness of ice bandings, it can also be illustrated by the speed-dependent packing density of particles. Under a small pulling speed, particles become close-packed ahead of the freezing interface, the proportion of the interstice water in the particle bandings is small and water transforms into segregated ice in large quantities. Thus the thickness of ice bandings is large when the pulling speed is small. Contrarily, under a large pulling speed, particles become loose-packed ahead of the freezing interface, the proportion of the interstice water in the particle bandings is large and so the thickness of ice bandings (i.e. segregated ices) is small. Therefore, the thickness of ice bandings decreases with increasing pulling speeds.

The packing density of particles, influenced by pulling speeds, determines the thickness of particle bandings and ice bandings. A promising mathematical model of the speed-dependent ice bandings may be proposed by considering the packing density of particles ahead of the freezing interface and the dynamic growth of the pore ice in future.

**Dynamic interface undercooling of ice bandings**

The dynamic growth of pore ice can be revealed by the interface undercooling.

In directional freezing colloidal suspensions, interface undercooling is related to the interface position within a linear thermal gradient [16, 24]. Here we obtained details of interface position and undercooling from Fig.1(a) and Movie S1. The dynamic growth process for an ice banding is shown as Fig.4. Figure 4 shows the evolution of freezing interface in a period of forming an ice banding. The freezing interface of supernatant is also presented to illustrate the interface undercooling. The interface undercooling $\Delta T$ can be calculated as $\Delta T=G\Delta x$ ($\Delta x$ is the interfacial separation of the suspension and its supernatant) with precision of 0.01K [16]. At the beginning of forming an ice banding, i.e., t=35s, a new ice lens sprouts near the interface position of the supernatant (red dotted line). Afterwards, the new ice lens grows up. However, the freezing interface of suspensions (blue dotted line) retreats to the cooling zone, indicating an undercooling compared with the interface position of the supernatant, as shown in Fig.4 (b), at t=110s. This undercooling will increase continuously and develop into a maximum until another new ice lens sprouts near the interface position of the supernatant at t=195s, as shown in Fig.4 (c).

The instantaneous interface position, speed and undercooling as functions of time are shown in Fig.5 (a), (b) and (c), respectively. Figure 5(a) shows the interface position with time extracted from Movie S1. The statistics of interface position are from photos of one per 25 seconds. The interface is not straight but migrates as a whole. So we chose half of the interface position to replace the movement of the whole interface. The data in Fig.5(a) are averages based on two independent statistics and the statistical errors are ±10% of the averages. Interface speed along with time in Fig.5 (b) is extrapolated from Fig.5 (a). We use a polynomial function to fit the data in Fig.5 (a). The differential of the polynomial function is Fig. 5 (b). Interface undercooling along with time is from the comparison between interface positions of the suspensions and its supernatant, as shown in Fig.5 (c). The data in Fig.5(c) are averages based on three independent statistics and the statistical errors are ±16% of the averages. From Fig.5 (a), the interface position exponentially retreats to the cold side until a new freezing interface appears close to the warm side. In Fig.5 (b), the interface speed exponentially decays in one period. Afterwards, it reduces to a minimum and starts a new period. It should be noted that the interface velocities at initial time are less than the pulling speed (V=0.8 μm/s) due to the resistance of the particles in front of the freezing interface. As for Fig.5 (c), the interface undercooling increases exponentially with time in one period, which means that the interface becomes undercooled with the dynamic accumulation of particles. Afterwards, it increases to a maximum and then starts a new period.

The whole processes of interface migration reveal the dynamic interface undercooling. Once a new ice appears in the beginning, the ice lens starts growing to keep up with the advancing isotherms. The growth of the freezing interface builds up a particle layer suggested in Fig. 3. The ice lens retreats to colder temperatures as particles accumulate and the interface growth is subjected to the increased resistance of the growing particle layer. The increasing resistance induces an enlarged interface curvature caused by the interactions between particles and the freezing interface. The curvature-dependent interface undercooling is consistent with the retreat of the interface position before the penetration of pore ices, as shown in Fig.5 (a). The maximum interface undercooling corresponds to the initial penetration of pore ice, indicated by the thickness of the particle bandings in a linear thermal gradient. Therefore, the dynamic interface undercooling is responsible for the formation of periodical ice bandings.

**Conclusions and Perspectives**

In this paper, we investigated speed-dependent ice bandings using an experimental facility with high precision to in-situ reveal the dynamic interface evolution of ice bandings. Two quantitative characteristics, particle packing density in front of the freezing interface and dynamic interface undercooling, were proposed to reveal speed-dependent ice bandings. The particle packing density ahead of the freezing interface decreases with increasing pulling speeds, which can be explained by the speed-dependent entropic ordering of particles. Through affecting the curvature undercooling of pore ices, the speed-dependent packing of particles can be used to explain speed-dependent thicknesses of ice bandings. Afterwards, the dynamic interface undercooling was investigated to reflect the curvature undercooling of pore ices. Quantitative details of transient interface positions, speeds and undercooling were given. All the evidences imply that the speed-dependent particle packing and dynamic interface undercooling are responsible for speed-dependent ice bandings.

In the future, a mathematical model should be proposed to describe the transient interface positions and speeds. Meanwhile, some more experiments are needed to in situ observe packing of particles in a three dimensional sample. In addition, some methods eliminating the ice bandings deserve further exploration and application in the cold regions [25].

**Acknowledgements**

J.Y. thanks M. Grae Worster for numerous discussions. This work is supported by National Natural Science Foundation of China (Grant Nos. 51571165 and 51701155), Free Research Fund of State Key Laboratory of Solidification Processing (100-QP-2014), the Fund of State Key Laboratory of Solidification Processing in NWPU (13-BZ-2014) and Innovation Foundation for Doctor Dissertation in NWPU (CX201703).

**List of figures**

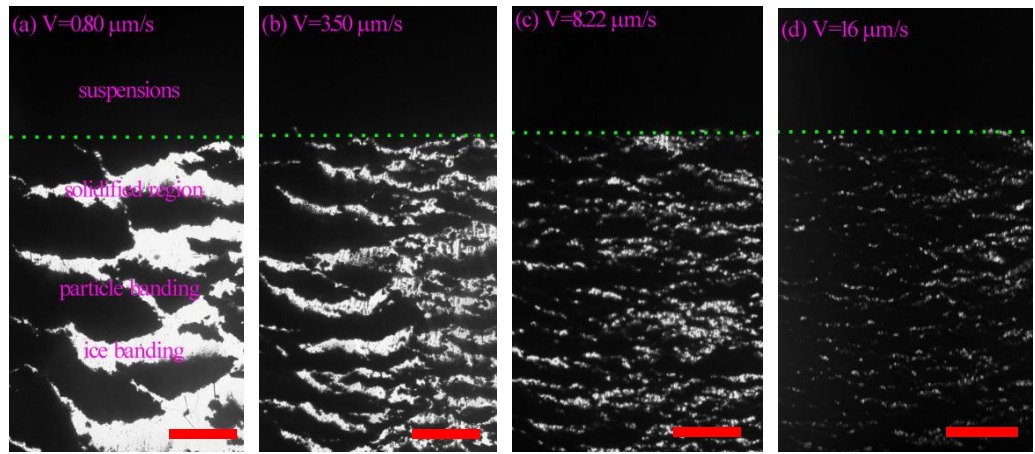

Fig.1 ice bandings with different pulling speeds. The bottom is the cooling zone and the top is a heating zone, building a linear thermal gradient G=7.23K/cm. Different pulling speeds are applied: (a) V=0.80 μm/s, (b) V=3.50 μm/s, (c) V=8.22 μm/s and (d) V=16 μm/s. Initial volume fractions of particles are $\phi_0$=15.97% and d=80 nm. The red scale bar is 200μm. The black part is particle banding and the white part is ice banding. The green dotted line is the freezing interface under which is the solidified region.

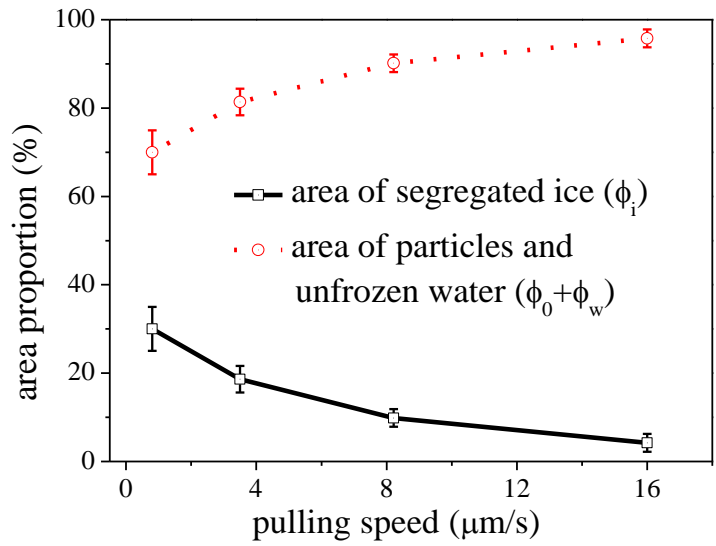

Fig.2 the volume fraction of segregated ice ($\phi_i$) and particle bandings ($\phi_0+\phi_w$) under different pulling speeds. The statistics are based on area fraction of segregated ice (i.e. ice banding) and particle bandings in the solidified region of Fig.1.

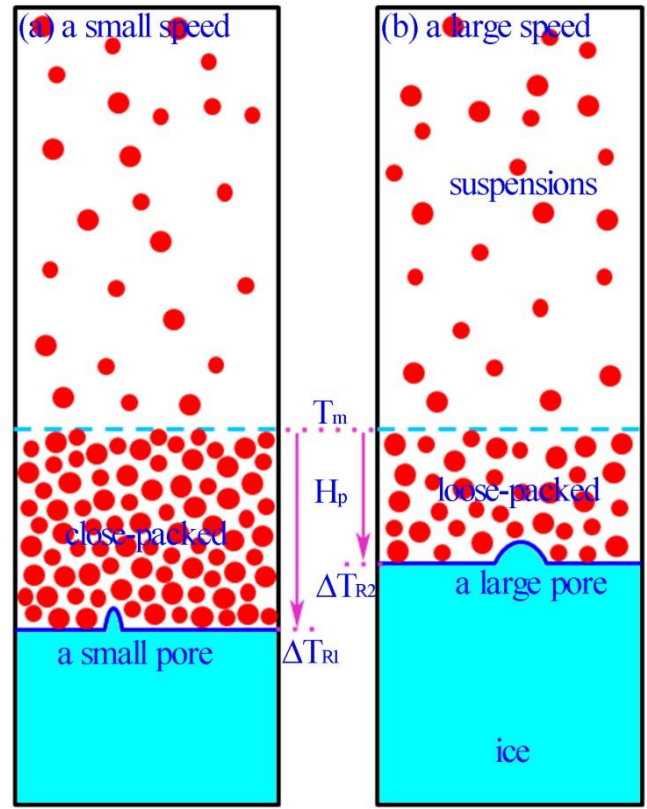

Fig.3 the schematic of speed-dependent particle packing and the dynamic growth of pore ice. $\Delta T_R$ is the curvature undercooling. $H_p$ is the thickness of particle bandings. When the pulling speed is small, the packing particles ahead of the freezing ice, become close-packed and hence a pore ice with a small radius can penetrate into the close-packed particles. Therefore, a large undercooling $\Delta T_{R1}$ is needed corresponding to a large thickness of particle bandings in a linear thermal gradient G, where $H_p=\Delta T_R/G$. On the contrary, a large pulling speed causes loose-packed particles, which implies a pore ice with a large radius, i.e. a small $\Delta T_{R2}$. The small undercooling corresponds to a small thickness of particle bandings. $T_m$ is the freezing point at which a new ice lens initiates.

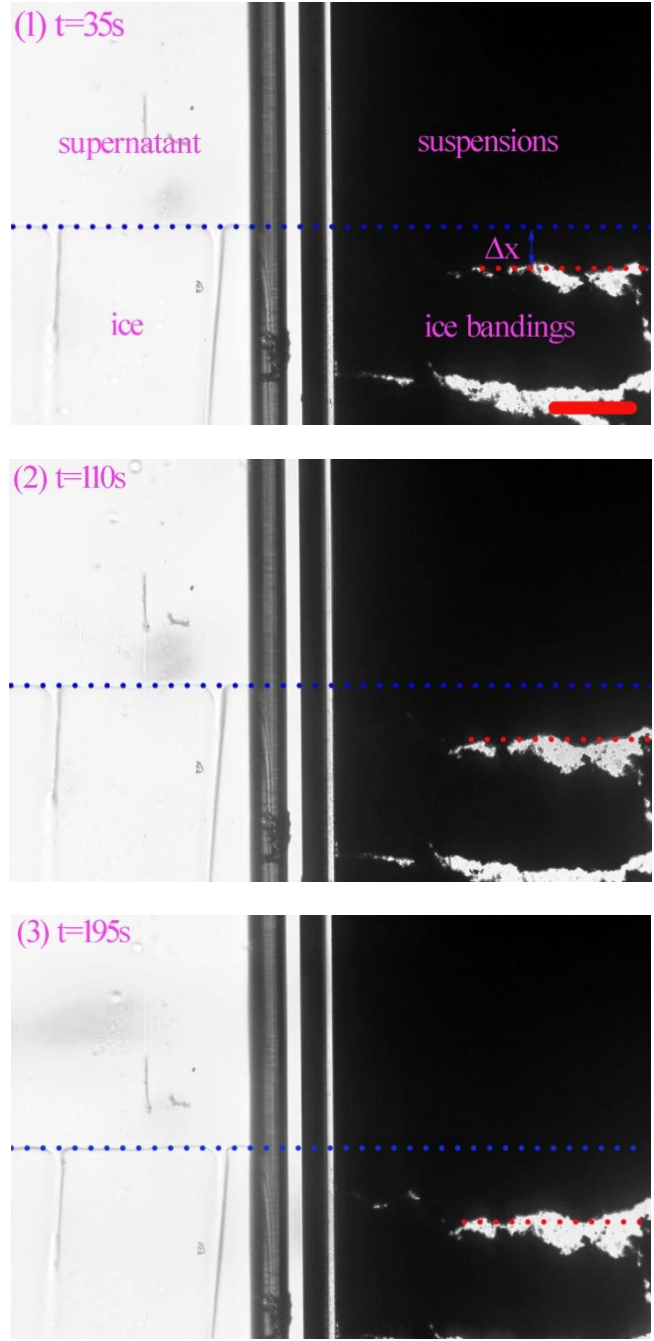

Fig.4 dynamic growth process of an ice banding. The blue dotted line is the interface position of supernatant and the red dotted line is the interface position of suspensions. Δx is the position discrepancy between the blue line and red line. Δx increases with time. The interface undercooling ΔT is calculated as ΔT=GΔx. The scale bar is 100 μm.

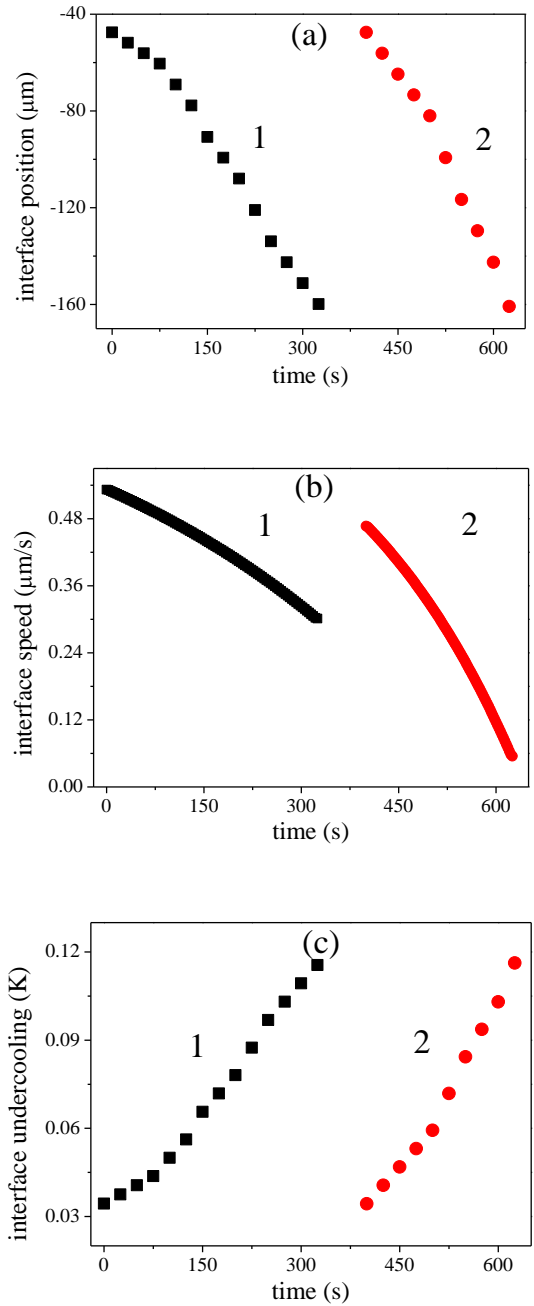

Fig.5 Freezing kinetics for ice bandings. (a) interface position with time extracted from Movie S1. The baseline is steady-state interface position of supernatants. The data in Fig.5(a) are averages based on two independent statistics and the statistical errors are ±10% of the averages. (b) interface speed with time extrapolated from Fig.5(a). We use a polynomial function to fit the data in Fig.5 (a). The differential of the polynomial function is Fig. 5 (b). (c) interface undercooling

with time from the comparison between interface positions of the suspension and its supernatant. The data in Fig.5(c) are averages based on three independent statistics and the statistical errors are ±16% of the averages. Two periods of forming ice bandings are given.